# Effects of High Intensity Ultrasound on BSCCO-2212 Superconductor

Tanya Prozorov[1,2], Brett McCarty[1], Zhihua Cai[1], Ruslan Prozorov[1§], and Kenneth S. Suslick[2]

[1]*Department of Physics & Astronomy and NanoCenter, University of South Carolina, Columbia, SC 29208*

[2]*School of Chemical Sciences, University of Illinois at Urbana-Champaign, Urbana, IL 61801*

Slurries of powdered $Bi_2Sr_2CaCu_2O_{8+x}$ superconductor in low volatility alkanes were treated with intense ultrasound. Significant enhancements of magnetic irreversibility as well as transport critical current are reported. The effects are dependent on the concentration of the slurry and are optimal for 1.5% wt. slurry loading. Electron microscopy shows that ultrasonic treatment leads to a change in grain morphology and intergrain welding. The observed enhancement of superconducting properties is consistent with the limitations in critical currents in BSCCO superconductor being due to intergrain coupling rather than intragrain pinning strength.

---

§ Corresponding author. prozorov@sc.edu. Tel: 803-777-8197. Fax: 803-777-3065

High transition temperature superconductors – $YBa_2Cu_3O_{7-x}$ (YBCO) and $Bi_2Sr_2CaCu_2O_{8+x}$ (BSCCO-2212) remain most studied and potentially useful members of cuprate family[1-5]. The bulk polycrystalline morphology of these ceramic superconductors is usually considered for large-scale practical applications (e.g., after being formed into wires). Therefore, significant effort has been devoted to improving their magnetic irreversibility and transport critical currents[3]. It is generally agreed that in YBCO these properties are mostly determined by the intragrain pinning parameters[5]. In BSCCO, on the other hand, intergrain coupling limits supercurrent flow[4]. New methods are therefore needed to enhance intergrain coupling in BSCCO and improve its irreversible magnetic properties. In this Letter, we report sonochemical modification of grain morphology and resulting improvement of the intergrain coupling in polycrystalline BSCCO-2212 superconductor. It was found that for moderate slurry loading, irreversible magnetic and transport properties are significantly enhanced.

Recently, we developed a method of sonochemical modification of granular $MgB_2$ superconductor[6]. In liquid-powder slurries irradiated with high-intensity ultrasound, acoustic cavitation induces turbulent flow and shock waves. Implosive collapse of bubbles during cavitation results in extremely high local temperatures, ~ 5000 K, and pressures, ~300 Mpa[7-9], and the shockwaves launched into the liquid create high-velocity collisions between suspended solid particles[6,10-12]. The estimated speed of colliding particles approaches half of the speed of sound in liquid. Effective temperatures at the point of impact can easily exceed melting temperature of most materials. Ultrasound-induced interparticle collisions are therefore capable of producing localized inter-particle melting and "neck" formation between particles[6,10-12]. This leads to significant improvement of intergrain coupling. The method has been successfully tested on $MgB_2$ superconductor. The irreversible magnetic properties of $MgB_2$, however, are mostly determined by *intra*grain pinning forces and the effect of ultrasound was to improve its reversible properties. Polycrystalline $Bi_2Sr_2CaCu_2O_{8+x}$ (325 mesh, <44 μm diameter; Alfa Aesar) was ultrasonically irradiated in 20 mL of decane for 120 min at 263 K under an argon flow (20 mL/min) using a direct-immersion ultrasonic horn (Sonics & Materials VCX-750 at 20 kHz, ~50 W/cm$^2$). Slurry loadings in the range of 0.5 to 10 %wt. were used. Resulting ultrasonically treated powders were collected by filtration, washed with dry pentane (30 mL×5), and air-dried overnight. Dry powders were pelletized at room temperature at a pressure of 2 GPa for 24 hours with an average sample mass of ~70 mg. Subsequent sintering was performed in a box furnace at 850 $^oC$ in air for 48 hours, followed by the quenching to room temperature. In order to compare our observations with other studies, we have used here a standard annealing procedure. We note in passing, however, that the properties of the sonicated powders can be even further optimized by changes in their handling (e.g., longer sintering times).

Scanning electron micrographs (SEM) were taken on a Hitachi S-4700 instrument. Samples were additionally characterized by powder X-ray diffraction and localized Energy-Dispersive X-ray spectroscopy. SEM images of $Bi_2Sr_2CaCu_2O_{8+x}$ powder after ultrasonic irradiation (Figure 1 (B)) show improved intergrain fusion, grain rounding, and greatly extended interconnection, compared to the starting material (Figure 1 (A)). These morphological changes are caused by partial melting of individual grains upon interparticle collisions. The observed modifications of morphology are consistent with the changes in $MgB_2$ reported earlier.[6]

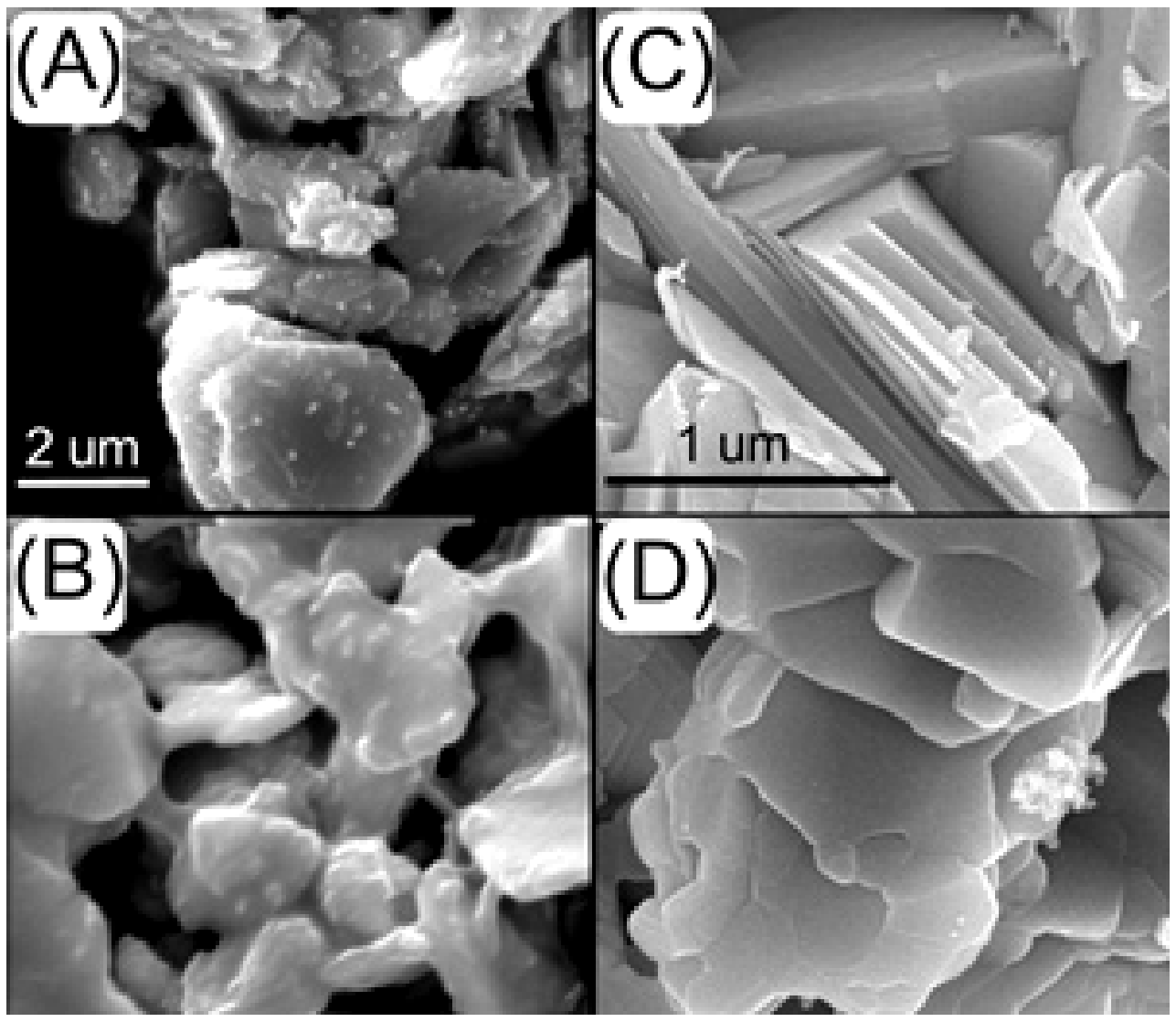

**Figure 1.** Scanning electron microscope images of (A) initial powder; (B) dried powder after sonication of 2% wt slurry loading; (C) sintered pellet made from initial powder; (D) sintered pellet made from sonicated material.

Transport measurements were performed by using the standard 4-point technique. Resistivity curves, normalized by the room temperature values, are shown in Figure 2. Importantly, the onset of superconductivity and steepness of the transition in sonicated BSCCO remained practically unchanged, indicating that observed morphological changes did not affect bulk chemical composition of initial polycrystalline material.

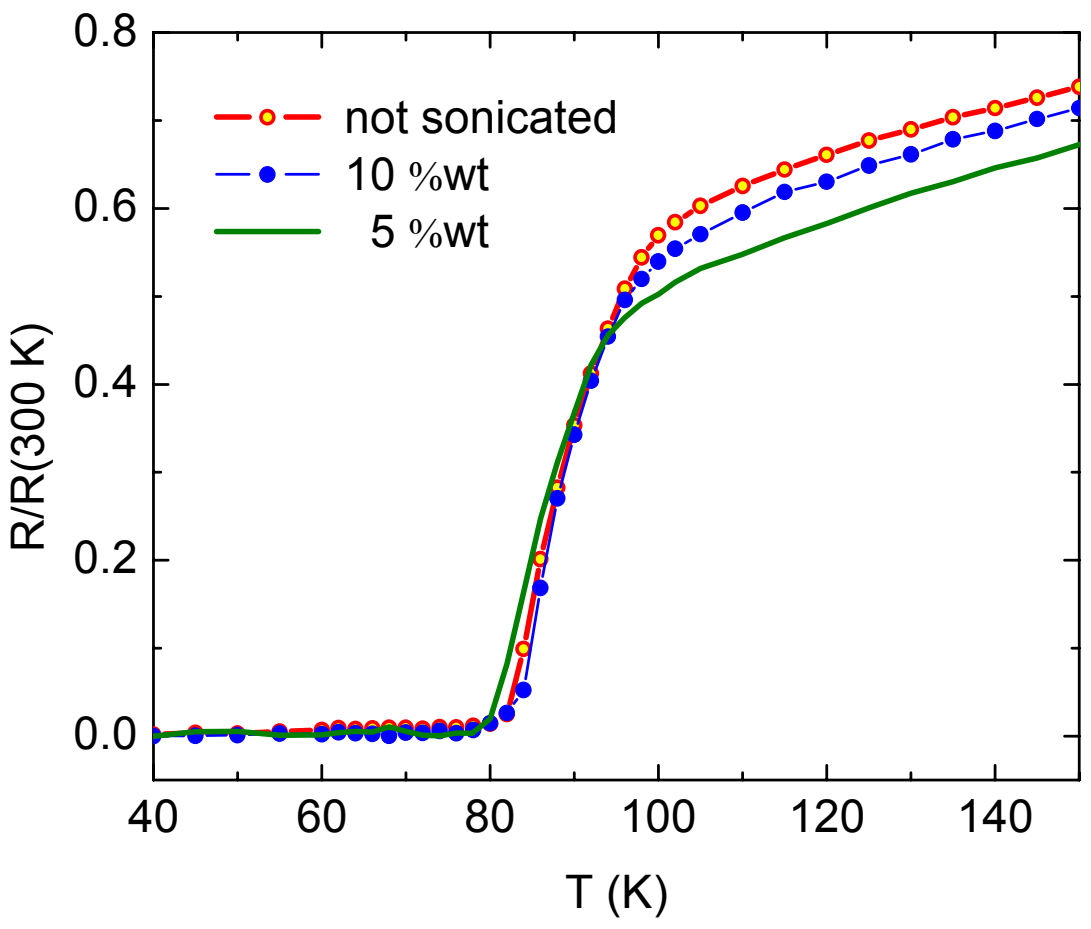

**Figure 2.** Resistance superconducting transition curves normalized by the room temperature value as a function of % wt. slurry loading during sonication.

Magnetic measurements were conducted by using Quantum Design MPMS magnetometer. For magnetic and transport measurements, samples were cut into slabs from the central part of sintered pellets, with all at similar sizes, $\sim 5\times 2\times 0.5$ mm. All reported measurements were performed with the magnetic field oriented along the long side of a slab in order to minimize the influence of demagnetization. To avoid uncertainty in sample comparison, magnetization curves were normalized by the initial slope, $\left|dM/dH\right|_{H\to 0} = V\left(4\pi(1-N)\right)^{-1}$ where $N$ is the demagnetization factor and $V$ is the sample volume. At higher magnetic fields the material is characterized by some magnetic susceptibility $\chi(H)$, and total measured magnetic moment is given by $M = \chi HV/(1-N)$. The volume of a superconducting fraction and the demagnetization factor are known only approximately. Normalization by the initial slope gives $4\pi M(1-N)/V = 4\pi\chi H = 4\pi m$. This has the clear meaning of volume magnetization expressed in gauss – a quantity independent of *V* and *N*, and representing behavior of an infinite slab with ideal initial susceptibility $4\pi\chi_{H\to 0} = -1$. We stress, however, that since our samples had very similar dimensions (and therefore, very similar volumes and demagnetization factors), raw data show the same trends as reported for normalized quantities. The normalization was used to extract critical current densities using Bean model for a long slab of $w\times b$ cross-section, $j_c\left[\mathrm{A\cdot cm^{-2}}\right] = 40m[\mathrm{G}]/w[\mathrm{cm}]/(1-w/3b)$, where $b \ge w$.

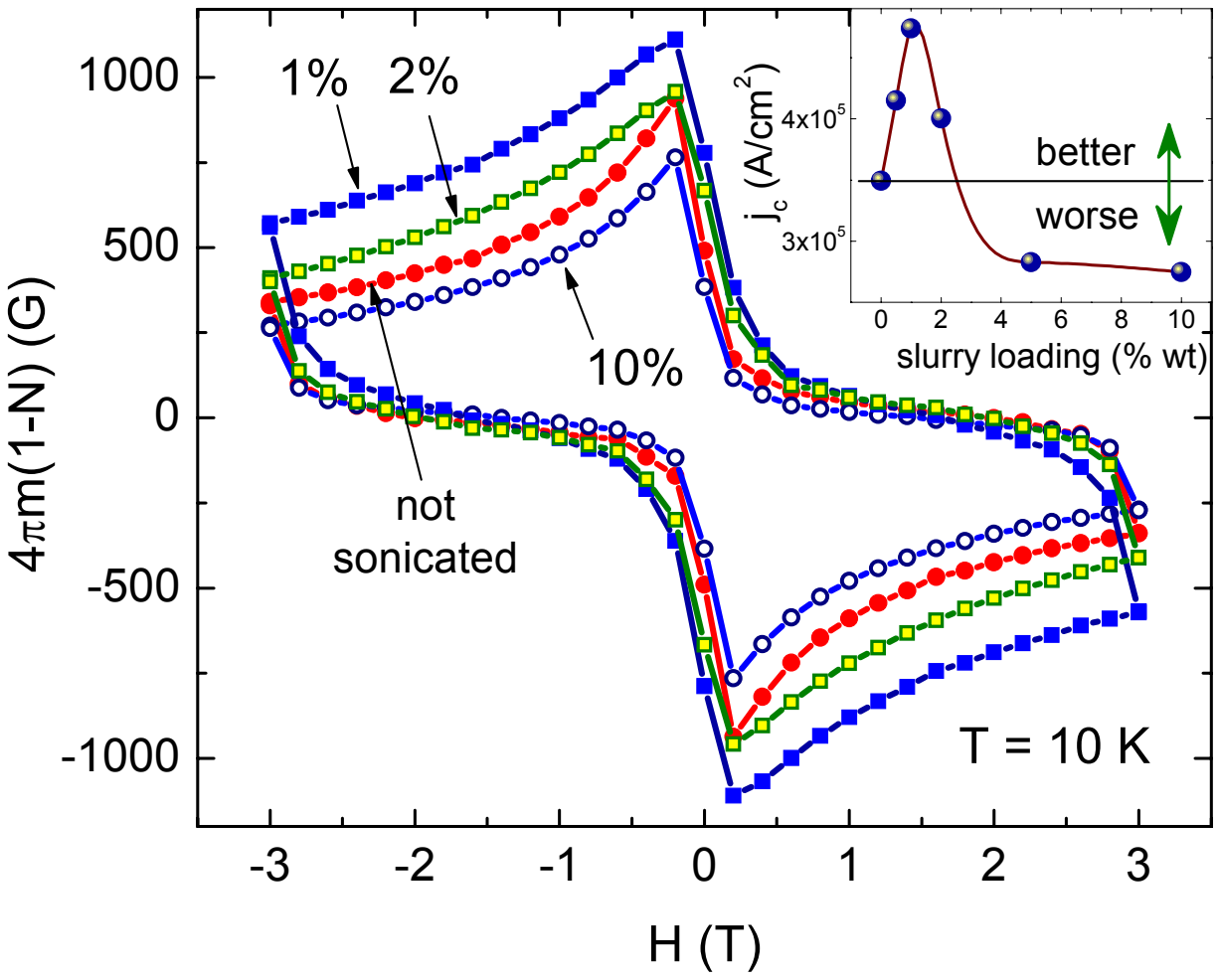

**Figure 3.** Effect of ultrasonic treatment on BSCCO slurries of various loading, indicated by %wt and arrows. *Inset* shows critical current density calculated at H=2000 Oe. See text for details.

Figure 3 shows normalized magnetization loops measured for samples of different slurry loading and for the reference sample. There is a clear enhancement of magnetic irreversibility in the entire range of magnetic fields for moderate slurry loadings. For large slurry loading, however, properties deteriorated. This is as expected, since large slurry loadings lead to inefficient low-speed collisions, which only result in destruction of grains surfaces. The inset shows critical current density evaluated from normalized magnetization at 2000 Oe. There is a significant enhancement of the critical current density (~40%) for sonication of 1% slurries compared to the non-sonicated, initial material.

In another set of experiments, magnetization was measured in a fixed 10 kOe magnetic field at various temperatures. To obtain each point, magnetic field was ramped up to 30 kOe and then reduced to 10 kOe to measure magnetization on a descending branch of the $M_{\downarrow}\left(H=10\,\text{kOe}\right)$ curve. Then the magnetic field was ramped down to a negative 30 kOe and increased back to +10 kOe, where $M_{\uparrow}\left(H=10\,\text{kOe}\right)$ was then measured. This procedure was repeated at each temperature shown in Figure 4.

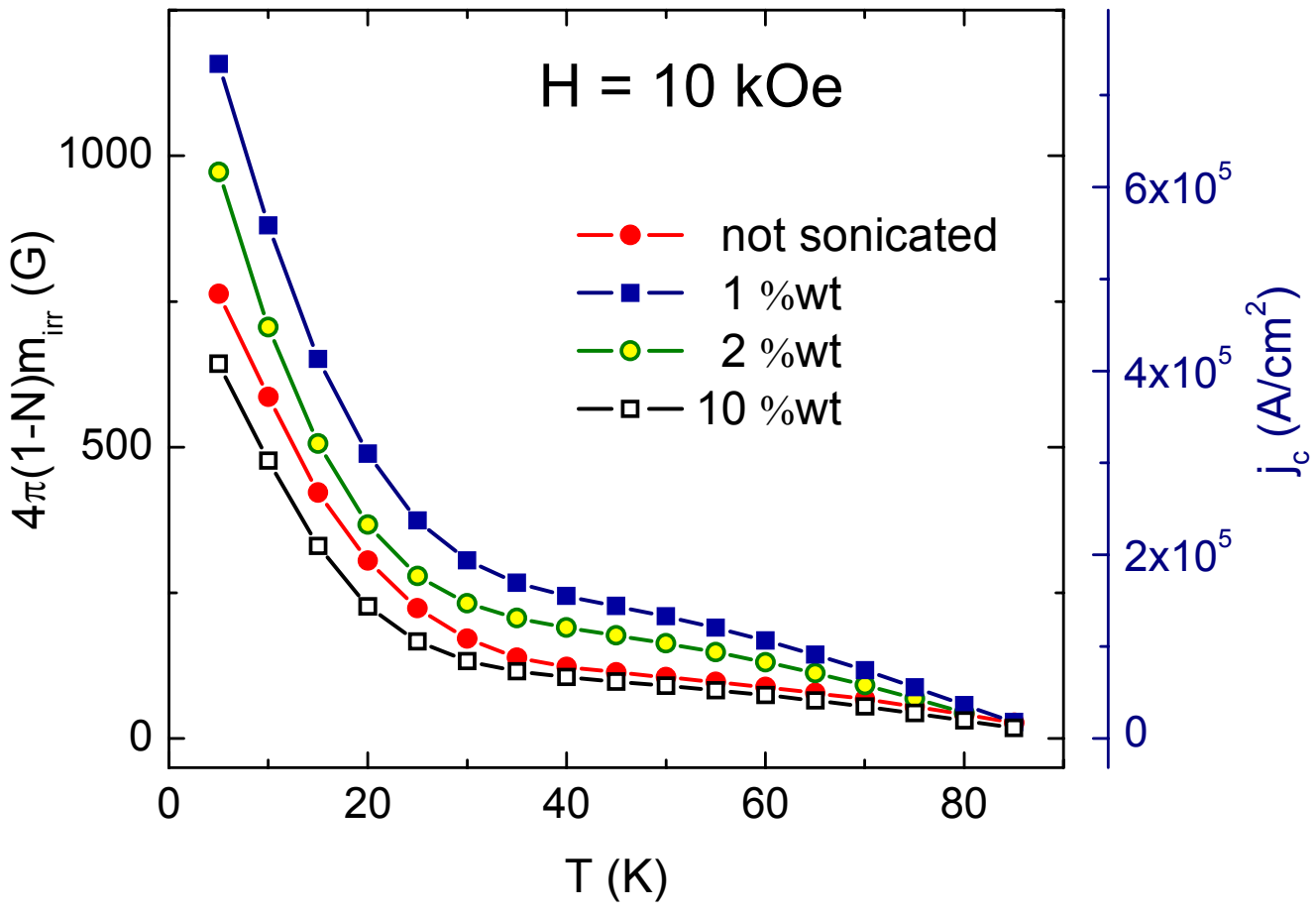

**Figure 4.** Irreversible magnetization measured at 10 kOe at various temperatures for samples of different slurry loading as described in the text. Right axis – critical current density evaluated by using Bean model.

The advantage of using this method is the ability to estimate irreversible part of magnetization, $M_{irr}=\left(M_{\uparrow}-M_{\downarrow}\right)/2$. This is a reliable way to estimate the critical current density from irreversible magnetization. Results, shown in Figure 4, are consistent with Figure 3 and indicate that observed enhancement of magnetic irreversibility persists over the entire temperature range. Moreover, the enhancement of critical current density at 77 K is about 60%: 55 kA/cm$^2$ for sonochemically treated sample versus 35 kA/cm$^2$ for non-treated sample.

In order to convincingly demonstrate the enhancement of true critical current, transport properties must be examined. The comparison of voltage-current characteristics measured in reference and sonicated samples is shown in Figure 5. There is a clear enhancement of the threshold value of the applied current. Since thick samples were used (the same as for the magnetic measurements), the absolute value for the critical current density is impossible to estimate due to unknown current density distribution throughout the sample. Nevertheless, the ratio of the threshold currents is in a good agreement with the magnetization data, with an enhancement of about 60%.

In conclusion, it is demonstrated that high intensity ultrasonic treatment of slurries of the high-$T_c$ superconductor BSCCO-2212 produces material with enhanced pinning properties. This method is effective for moderate slurry loadings. Our results outline a generalizable new direction for the improvement of vortex pinning in ceramic superconductors.

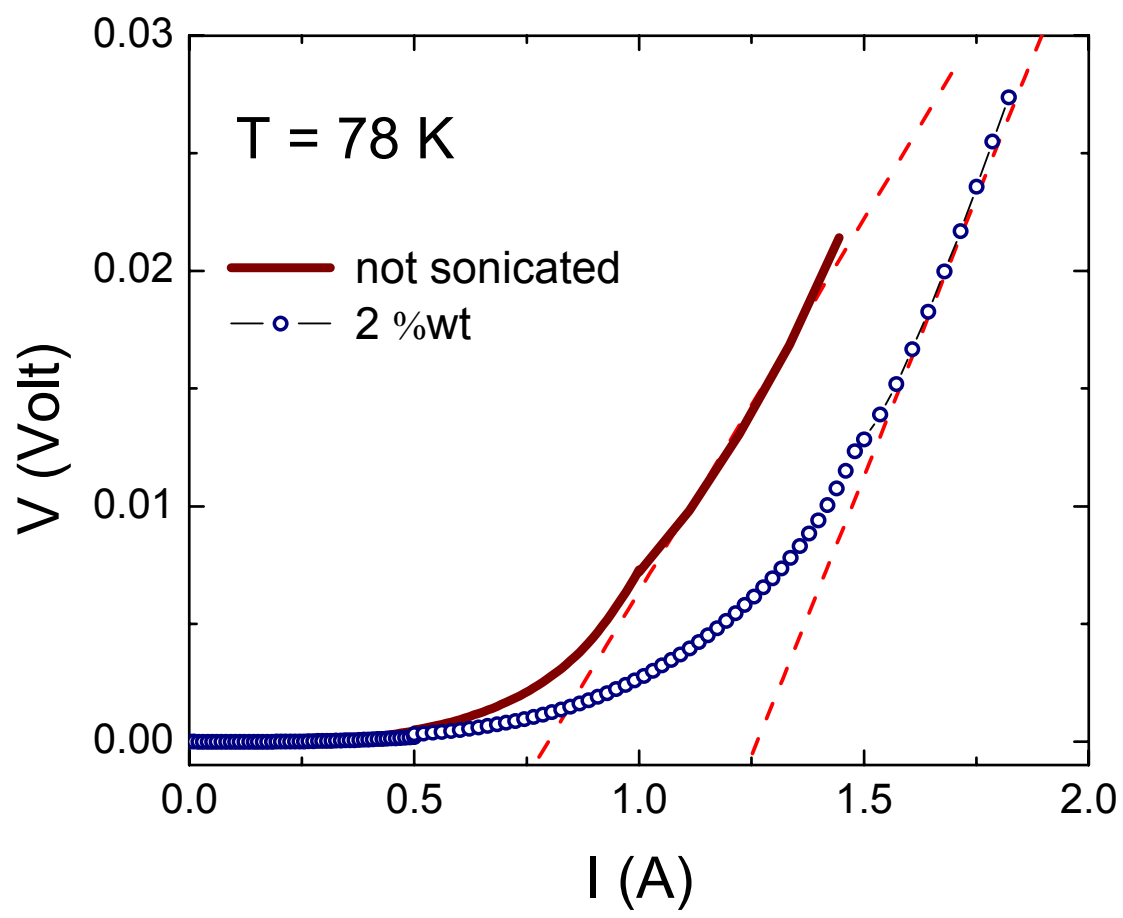

**Figure 5.** V-I characteristics measured in reference (left line) and an ultrasonically-treated sample (sonicated as a 2% wt. slurry). The enhancement of critical current is estimated about 60 %.

Discussions with V. Geshkenbein, A. Gurevich, A. Polyanskii, and V. Vlasko-Vlasov are greatly appreciated. This work was supported by the NSF-EPSCoR Grant EPS-0296165, a grant from the University of South Carolina Research and Productive Scholarship Fund, and the donors of the American Chemical Society Petroleum Research Fund. The SEM study was carried out in the Center for Microanalysis of Materials (UIUC), which is partially supported by the DOE under Grant DEFGO2-91-ER45439.

## REFFERENCES

[1] R. Funahashi, Int. J. of Cond. Matt. Res. **6** (1), 49 (2001).

[2] R. Flukiger, G. Grasso, J. C. Grivel, F. Marti, M. Dhalle, and Y. Huang, Supercon. Sci. Tech. **10** (7A), A68 (1997).

[3] http://www.asc.wisc.edu/bscco/bscco.htm.

[4] D. C. Larbalestier, IEEE Trans. 7, 90 (1997).

[5] D. M. Feldmann, D. C. Larbalestier, D. T. Verebelyi, W. Zhang, Q. Li, G. N. Riley, R. Feenstra, A. Goyal, D. F. Lee, M. Paranthaman, D. M. Kroeger, and D. K. Christen, Appl. Phys. Lett. **79**, 3998 (2001).

[6] T. Prozorov, R. Prozorov, A. Snezhko, and K. S. Suslick, Appl. Phys. Lett. **83**, 2019 (2003).

[7] K. S. Suslick, D. A. Hammerton, and R. E. Cline, Jr., J. Amer. Chem. Soc. **108**, 5641 (1986).

[8] W. B. McNamara III, Y. Didenko, K. S. Suslick, Nature **401**, 772 (1999).

[9] W. B. McNamara III, Y. Didenko, K. S. Suslick, J. Phys. Chem. B **107**, 7303 (2003).

[10] K. S. Suslick and S. J. Doctycz, Science **247**, 1067 (1990).

[11] K. S. Suslick and G. J. Price, J. Ann. Rev. Mat. Sci. **29**, 295 (1999).

[12] T. Prozorov, R. Prozorov, and K. S. Suslick, J. Amer. Chem. Soc. (Comm.) **in press** (2004).